\documentclass[review]{elsarticle}

\usepackage{lineno,hyperref}
\newcommand{\bhline}[1]{\noalign{\hrule height #1}}

\journal{Biological Psychology}

\begin{document}

\begin{frontmatter}

\title{Economical Visual Attention Test for Elderly Drivers}

\author[mymainaddress,mysecondaryaddress]{Akinari Onishi \corref{mycorrespondingauthor}}
\cortext[mycorrespondingauthor]{Corresponding author}
\ead[url]{http://onishi.starfree.jp/}
\ead{onishi-a@es.kagawa-nct.ac.jp}

\address[mymainaddress]{Chiba University, 1-33, Yayoicho, Inage-ku, Chiba-shi, Chiba, Japan}
\address[mysecondaryaddress]{National Institute of Technology, Kagawa College, 551 Kohda, Takuma-cho, Mitoyo-shi, Kagawa, Japan}

\begin{abstract}
Traffic accidents involving elderly drivers are an issue in a super-aging society. 
A quick and low-cost aptitude test is required to reduce the number of traffic accidents. 
This study proposed an oddball-serial visual search task that assesses the individual's performance 
by his or her responses to the presence of cued stimuli on the screen. 
Task difficulty varied by changing the number of simultaneous stimuli; 
Accordingly, low performers were detected. 
In addition, performance correlated with age. 
This implies that individual characteristics related to driving performance that decline with age 
can be detected by the proposed task. 
Since the task requires low-cost devices (computer and response button), 
it is feasible for use as a quick and low-cost aptitude test for elderly drivers. 
\end{abstract}

\begin{keyword}
\texttt{Elderly drivers\sep aptitude test\sep oddball\sep visual search\sep EEG}
\MSC[2010] 00-01\sep  99-00
\end{keyword}

\end{frontmatter}

%\linenumbers

\section{Introduction}
It was estimated in 2001 that approximately a quarter of the people in the Organisation for 
Economic Co-operation and Development (OECD) Member countries would be older than 65 years by 2050 \cite{oecd2001ageing}. 
Japan became a super-aging society in 2007, and numerous associated issues have arisen \cite{muramatsu2011japan}. 
One issue in a super-aging society is traffic accidents involving elderly drivers \cite{nakano2008method}. 

Elderly drivers have unique driving characteristics. 
For example, elderly drivers are relatively likely to crash at intersections \cite{mayhew2006collisions}.
Such crashes are related to failures of attention \cite{caird2005older}. 
In fact, traffic accidents involving elderly drivers are related to 
their physical, motor, sensory, and cognitive factors. 
Since visual function changes with aging, 
various studies of vision and vision-related cognitive functions have been conducted. 
The correlation between vision tests and crash records was weak, 
while attention scores were correlated with crash risk and driving performance \cite{anstey2005cognitive}. 
To prevent serious traffic accidents involving elderly drivers, assessments of visual attention are essential. 
Such assessments would enable elderly drivers to reduce risks by, for example, 
installing driver assistance systems \cite{bengler2014three}, or by training physical or cognitive functions \cite{hiraoka2016cognitive}.  

Various driving assessment systems utilize driving simulators (DSs). 
Most DS utilize 
driving scenes that contain events, such as a pedestrian crossing in front of the driving vehicle.
A study employed a DS to investigate the relationship between mental workload and age \cite{michaels2017driving}. 
In addition, some DS tests have been proposed to measure the characteristics of elderly drivers. 
For example, one DS test measures the range of the peripheral visual area 
that is related to the cognitive function named the useful field of view (UFOV) \cite{nakano2008method}. 
Furthermore, a DS test has been reported that clarifies the characteristics of the elderly driver's response 
by recording their palmar sweating response, galvanic skin response, and driving skills \cite{takahashi2017driving}. 
Therefore, DS tests are helpful for revealing the characteristics of elderly drivers in detail. 
However, these are not suitable as widely applied aptitude tests for elderly drivers 
because of test-related costs and high time demands. 
For example, most DSs are expensive and physically large.
Moreover scenario-based tests should be repeated many times in order to permit robust statistical analysis. 
In a super-aging society, 
many elderly drivers require aptitude tests. 
Thus, tests should be able to be conducted given limited time, space, and cost. 
Accordingly, minimal psychological tests should be considered.

Numerous psychological and physiological measures have been used 
to evaluate the driving performance of elderly drivers \cite{papantoniou2017review}.
Cognitive test scores (e.g., Mini-Mental State Exam) have been shown to correlate with in-traffic driving performance scores \cite{odenheimer1994performance}.
Moreover, driving relies on vision, including visual attention \cite{owsley2010vision}, and 
visual cognitive function is influenced by aging \cite{owsley1991visual,ishimatsu2010age}.
One well-studied visual function score is the UFOV \cite{miura1986coping, ball1993useful}. 
The UFOV declines with age \cite{b2000effects, lunsman2008predicts, seiple1996age} and is highly predictive of the likelihood of accidents among elderly drivers \cite{ball1993visual}.
The UFOV is measured using a salient stimulus in peripheral vision, in a form of pop-up task. 
However, dangers hidden in actual scenes are not salient.  
In addition to those subjective evaluation, performance should be estimated by 
an objective evaluation using physiological signals such as electroencephalography (EEG) \cite{wester2008event}.

This study developed a low-cost, rapid aptitude test for elderly drivers that combined a visual search task with an oddball task. 
Since psychological tests are more economical than DS tests, 
this study focused on a simple psychological test that required only a computer and a response button. 
The difficulty of the task was varied in three conditions 
to clarify each subject's performance.
Young and elderly participants were recruited to determine 
individual differences in each group, in addition to differences between groups.

\section{Material and methods}

\subsection{Participants}
% Basic information
Fifteen elderly participants 
(68.7$\pm$3.0 years, 8 male, with a valid driver's license) 
and 10 young participants 
(23.1$\pm$2.9 years, 8 male) participated in this study. 
All participants provided written informed consent before the experiment began.
This study was approved by the Internal Ethics Committee at Chiba University. 
All experiments were conducted in accordance with the Declaration of Helsinki. 
Profiles of participants are summarized in Table.~\ref{Table:profile}. 

% Participant information
\begin{table}[!t]
\caption{Profiles of participants.}
\label{Table:profile}
\begin{center}
\begin{tabular}{c|cccccc}
\bhline{2pt}  
ID 	& Age 	& Gender 	& Handedness 	& Vision Correction	& Driver's licence (experience) 	\\
\hline
y1	& 31	& M			& Right		& Spectacles			& Yes (13 years)					\\
y2	& 22	& M			& Right		& Contact lenses	& Yes (3 years)						\\
y3	& 23	& M			& Right		& Spectacles			& Yes (5 years)						\\
y4	& 22	& M			& Right		& No				& Yes (4 years)						\\
y5	& 23	& M			& Right		& Contact lenses	& Yes (4 years)						\\
y6	& 21	& M			& Right		& Contact lenses	& Yes (3 years)						\\
y7	& 24	& M			& Right		& Contact lenses	& Yes (4 years)						\\
y8	& 22	& M			& Right		& Contact lenses	& No								\\
y9	& 21	& F			& Right		& Contact lenses	& No								\\
y10	& 22	& F			& Right		& Spectacles			& Yes (3 years)						\\
e1	& 69	& M			& Right		& Spectacles			& Yes (42 years)					\\
e2	& 65	& M			& Right		& No				& Yes (28 years)					\\
e3	& 70	& M			& Right		& Spectacles			& Yes (51 years)					\\
e4	& 70	& M			& Right		& No				& Yes (29 years)					\\
e5	& 66	& F			& Right		& Spectacles			& Yes (34 years)					\\
e6	& 67	& F			& Right		& Spectacles			& Yes (46 years)					\\
e7	& 71	& F			& Right		& No				& Yes (53 years)					\\
e8	& 69	& M			& Right		& Spectacles			& Yes (46 years)					\\
e9	& 76	& M			& Right		& No				& Yes (55 years)					\\
e10	& 69	& M			& Both		& Spectacles			& Yes (36 years)					\\
e11	& 72	& M			& Right		& Spectacles			& Yes (54 years)					\\
e12	& 67	& F			& Right		& Spectacles			& Yes (51 years)					\\
e13	& 65	& F			& Right		& No				& Yes (26 years)					\\
e14	& 65	& F			& Right		& No				& Yes (47 years)					\\
e15	& 70	& F			& Right		& Spectacles			& Yes (37 years)					\\
\hline
\end{tabular}
\end{center}
\end{table}

\subsection{Oddball-serial visual search task}
\begin{figure}[!t]
\centering
\includegraphics[width=10cm]{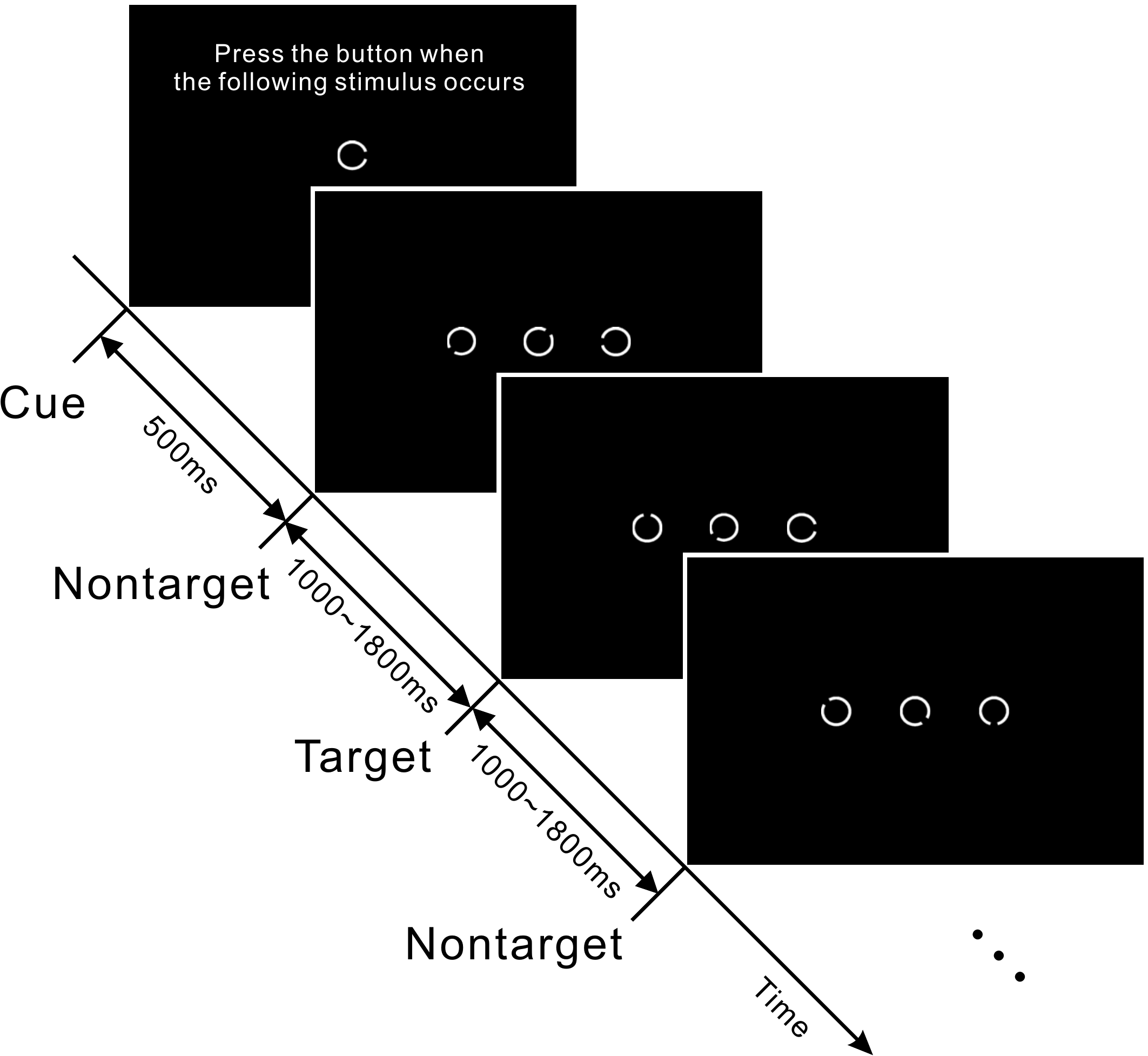}\\
\caption{Example of the oddball-serial visual search task. 
First, the target stimulus was cued. 
After 500 ms,  
three stimuli were simultaneously presented in a horizontal orientation 
in the middle of the screen (P3 condition). 
New sets of stimuli were continuously presented every 1000 to 1800 ms. 
The participant pushed a button using his or her right-hand 
as quickly as possible (right-hand condition) 
if the target stimulus was present anywhere on the screen. 
If no target stimulus appeared on the screen (nontarget stimulus), 
the participant was instructed to ignore the stimulus. 
The task was repeated during a session. 
}
\label{fig:OSVS}
\end{figure}

%Stimuli
In this study, a participant was asked to respond to the cued stimulus 
by pushing a button. 
The task was an oddball-serial visual search (OSVS). 
The participant was seated 60 cm 
in front of the monitor (E178FPc, DELL), 
holding a button box. 
This study employed 8 circular stimuli; 
part of the circle in one of 8 directions was missing. 
The stimuli are similar to Landolt Cs.
Figure~\ref{fig:OSVS} demonstrates an example of a trial. 
First, the participant remembered a cued target stimulus. 
Then, multiple stimuli were horizontally presented simultaneously for 500 ms. 
If the target stimulus appeared, 
the subject was required to press the button as quickly as possible. 
After a response was made, the next stimuli were presented after a 500 to 1300 ms delay; 
the stimulus-onset asynchrony was 1000 to 1800 ms.

%Figure of OSVS task

\begin{figure}[!t]
\centering
\includegraphics[width=10cm]{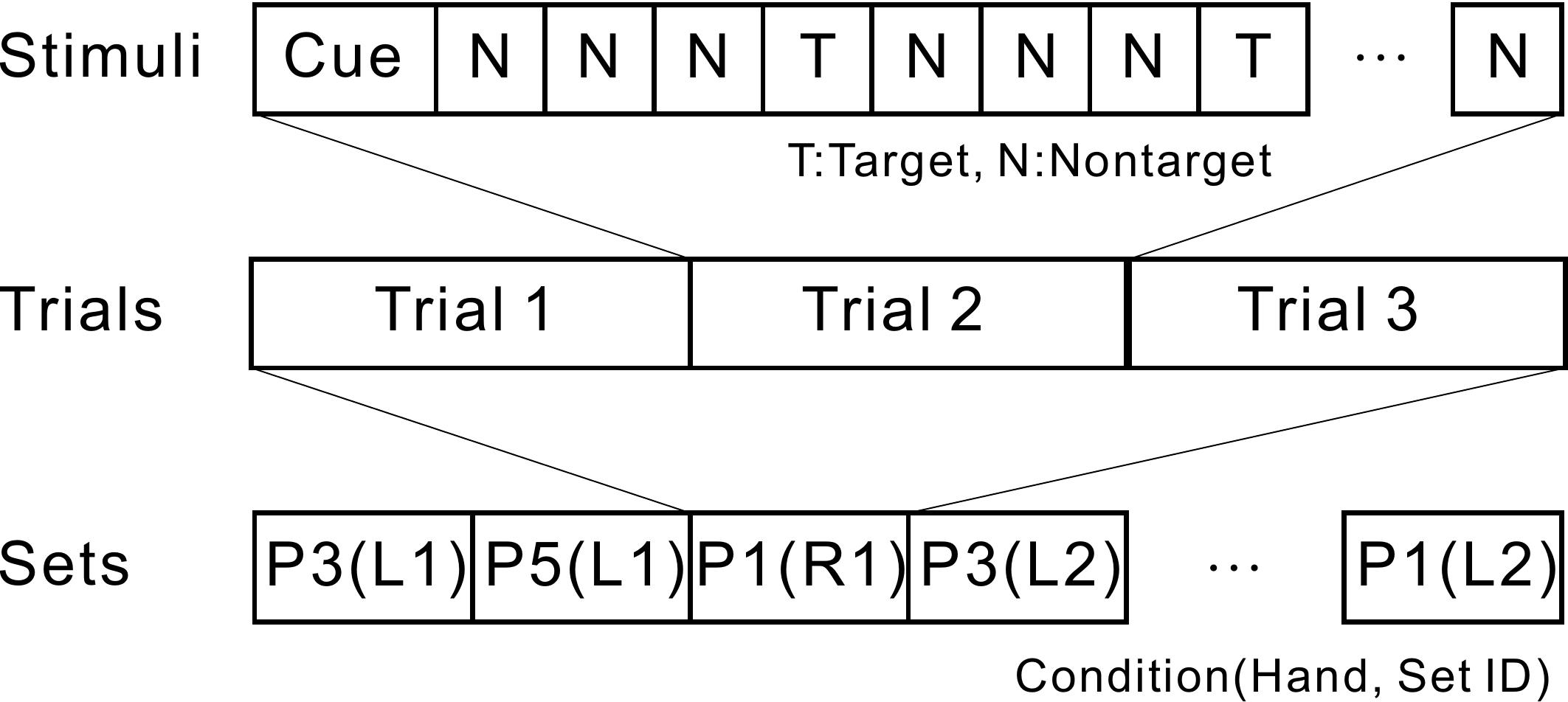}\\
\caption{Experimental procedure. 
Each trial began with a cue. 
After the cue, 40 stimuli were displayed pseudo-randomly, in succession. 
Two types of stimuli were provided: target stimuli (T), which required a response, 
and nontarget stimuli (N), which did not require a response. 
Three blocks comprised a set of experimental trials. 
Each set consisted of one experimental condition, namely 
the stimulus condition (P1, P3, P4), 
hand condition (R: right, L: left), and set ID (1, 2).
Each stimulus condition contains 480 stimuli. 
For example, P3 condition comprise of 4 sets, 
namely P3(R1), P3(R2), P3(L1), and P3(L2) sets.  
Those 4 sets (hands and twice) contain 3 trials, and each trial has 40 stimuli, respectively. 
}
\label{fig:Procedure}
\end{figure}

\begin{figure}[!ht]
\centering
\includegraphics[width=5cm]{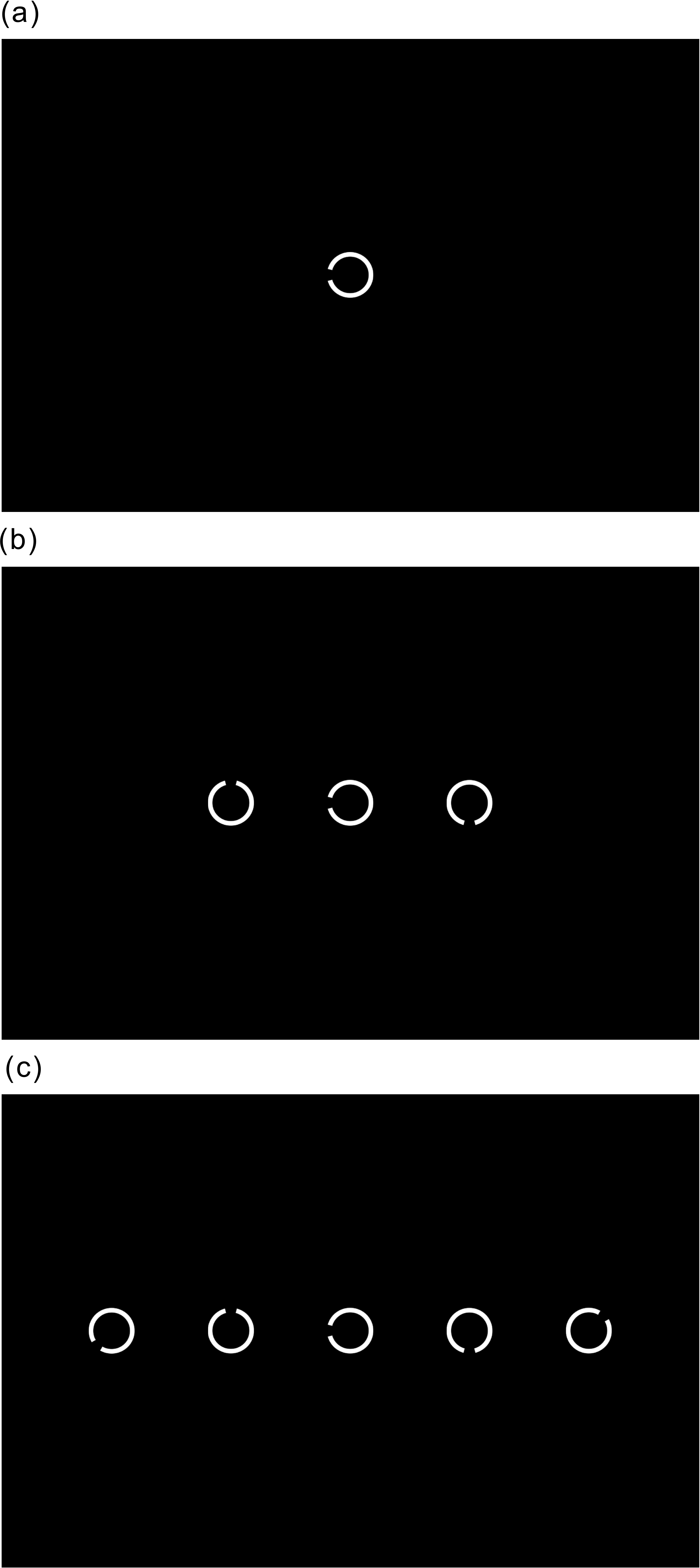}\\
\caption{Stimulus conditions. 
P1 condition (a) a single circle 
with a missing segment, displayed at the center of the screen. 
P3 (b) and P5 (c) conditions show three and five circles horizontally 
along the vertical center of the screen, respectively.
}
\label{fig:conditions}
\end{figure}

\begin{figure}[!t]
\centering
\includegraphics[width=8cm]{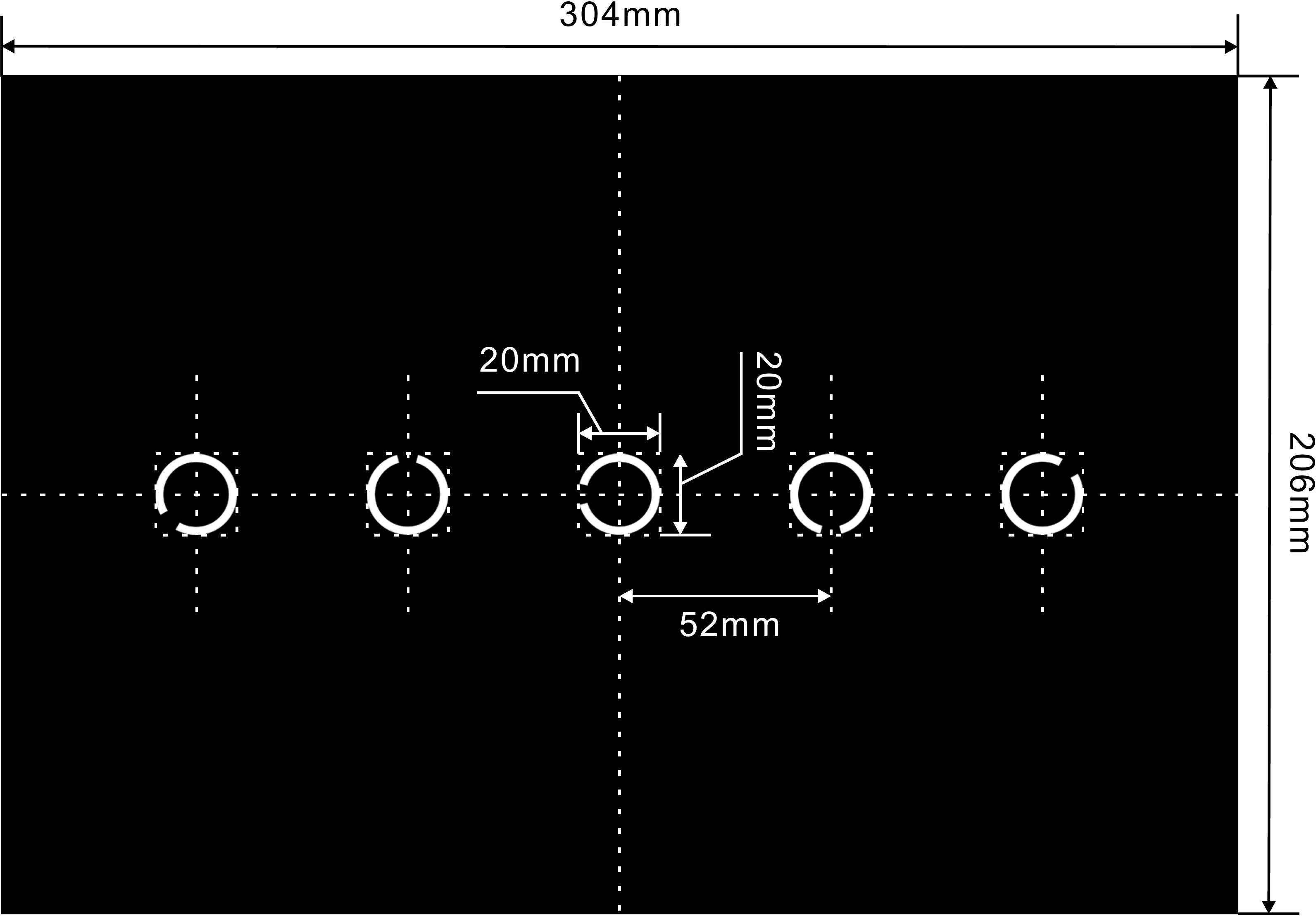}\\
\caption{Location and size of the stimuli on the monitor. 
}
\label{fig:Size}
\end{figure}

The experimental design is shown in Fig.~\ref{fig:Procedure}. 
There were 40 stimulus presentation in each block of trials, 
in which target stimuli appeared 8 times. 
That is, the probability that the target stimulus appeared was 1/5.
Three blocks of trials were conducted in an experimental sequence.
The sequence was conducted with right and left hand conditions. 
Three stimulus conditions were prepared; 
one (P1), three (P3), or five (P5) different stimuli were 
presented simultaneously (see Fig.~\ref{fig:conditions}). 
Furthermore, all conditions were repeated twice. 
There were 480 trials for each stimulus condition 
(40 stimuli $\times$  2 hands $\times$ 3 trials $\times$ 2 times), 
in which the target stimulus was presented 96 times. 
Note that the experimental order was pseudo-randomized. 
The duration of the experiment was approximately 2 hours. 
Participants rested for one minute between blocks of trials. 
The location and size of the stimuli are shown in Fig.~\ref{fig:Size}.

\subsection{EEG recording}

EEG signals were recorded during the experiment. 
This study used the Polymate mini AP108 (Miyuki Giken Co., Ltd, Tokyo, Japan), 
a portable EEG amplifier with active electrodes. 
Electrodes were located at Fpz, Fz, Cz, Pz, and Oz 
in accordance with the international 10-20 system. 
The reference and ground electrodes were placed at A2 and Afz, respectively. 
The sampling rate was 500 Hz. 
Filters applied consisted of a hardware low-pass filter (cut-off frequency: 30 Hz), 
high-pass filter (time-constant: 1.5 s), 
and notch filter (50 Hz). 
The sensitivity of the amplifier was 20 $\mu$V.

\subsection{Analysis}

This study focused on visualizing the raw data 
to confirm the presence of individual differences among the subjects. 
Task performance was analyzed using the 
frequency of true-positive (TP), true-negative (TN), 
false-positive (FP), and false-negative (FN) responses, 
in addition to accuracy, precision, and sensitivity.  
Further, descriptive statistics were used to summarize 
the results of the experiments. 
Medians were used as the measure of central tendency of the data, 
because of the presence of ceiling effects.
Correlation analysis was also used 
to clarify the relationships among profiles of participants 
and performance.

\clearpage

\section{Results}

\subsection{Individual performance}

\subsubsection{TP, TN, FP, FN}

\begin{figure}[!t]
\centering
\includegraphics[width=10cm]{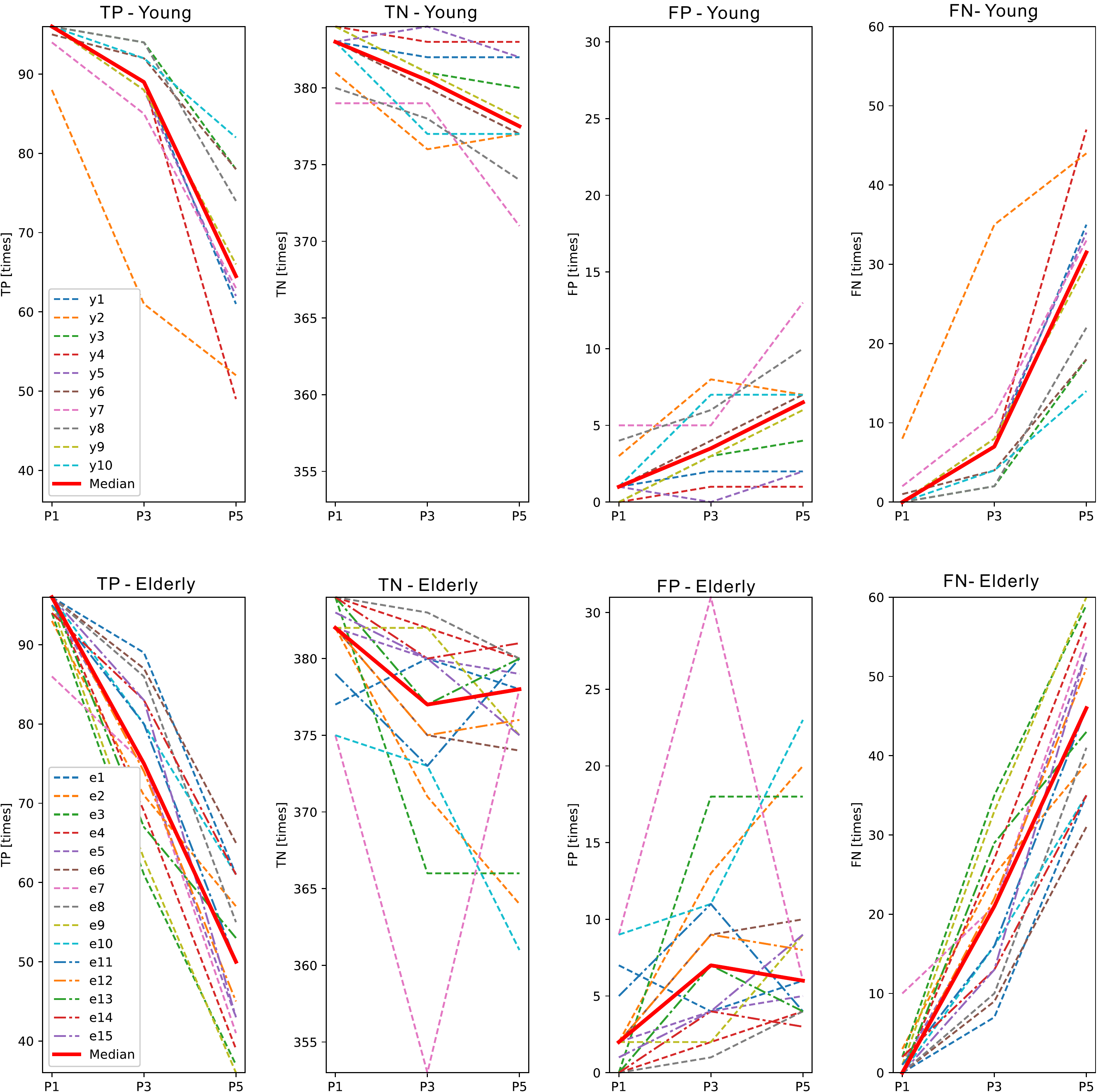}\\
\caption{Individual's TP, TN, FP, and FN responses in each stimulus condition (P1, P3, P5).}
\label{fig:conf_place}
\end{figure}

To clarify each individual's performance, 
all participants conducted the OSVS task 
for each stimulus condition (P1, P3, P5). 
Figure~\ref{fig:conf_place} represents 
the number of TP, TN, FP, and FN responses during the task. 

The median TP of the young participants was 96 in P1, 
89 in P3, and 64.5 in P5 (TP - Young). 
Median TP of the elderly participants was  
96 in P1, 
75 in P3, and 50 in P5 (TP - Elder). 
Large variance in TP can be seen in P3 and P5 for elderly participants, 
and in P5 for young participants. 
Participant y2 exhibited the lowest TP in P3, at 52 such responses, 
which was far from the median. 
The TP of participants y2 and y4 in P5 was 52 and 49, respectively. 
Participants e3, e4, and e9 produced 
37, 39, and 36 such responses, respectively. 
TP during P5 was low for most elderly participants. 

The variance in TN was small across conditions. 
TN for young participants was 383, 380.5, and 377.5 (TN - young), 
while TN for elderly participants was 382, 377, and 378 (TN - Elder) 
for P1, P3, and P5 conditions, respectively.
TN of participant e7 in P3 condition was 353, 
which was lower than the median. 

The line chart shows that the FP and FN plots are essentially inverted versions of the TN and TP plots, respectively.

\subsubsection{Accuracy, precision, and sensitivity}

Accuracy, precision, and sensitivity were calculated, as indicated in Fig.~\ref{fig:acc_place}. 
Median accuracy of the young participants was 99.8, 98.2, and 92.5\%, 
while that of elderly participants was 99.6, 94.4, and 87.9\% for P1, P3, and P5 conditions, respectively. 
Participant y2 exhibited the lowest accuracy among young subjects for all stimulus conditions. 
Participants e3 and e7 exhibited decreases in the P3 condition. 
Most elderly subjects demonstrated low accuracy in the P5 condition. 

The precision values reflected individual differences among the elderly participants. 
Median precision of the young participants was 99.0, 96.3, and 91.9\%, 
while that of elderly participants was 98.0, 90.6, and 87.2\% for P1, P3, and P5 conditions, respectively.  
Some elderly participants demonstrated high precision in P3 and P5 conditions, 
while participants e2, e3, and e10 exhibited marked decreases in P3 and P5 conditions. 
These tendencies were not observed in accuracy and sensitivity. 
Participant e7 exhibited the lowest precision in P3, but approximately median precision in P5. 

Although the shape of the trend in median sensitivity was similar to that of accuracy, 
the trends for individual participants differed between accuracy and sensitivity. 
For example, accuracy of participant e4 was close to the median, while sensitivity was among the worst.

\subsubsection{Response time}
\begin{figure}[!t]
\centering
\includegraphics[width=10cm]{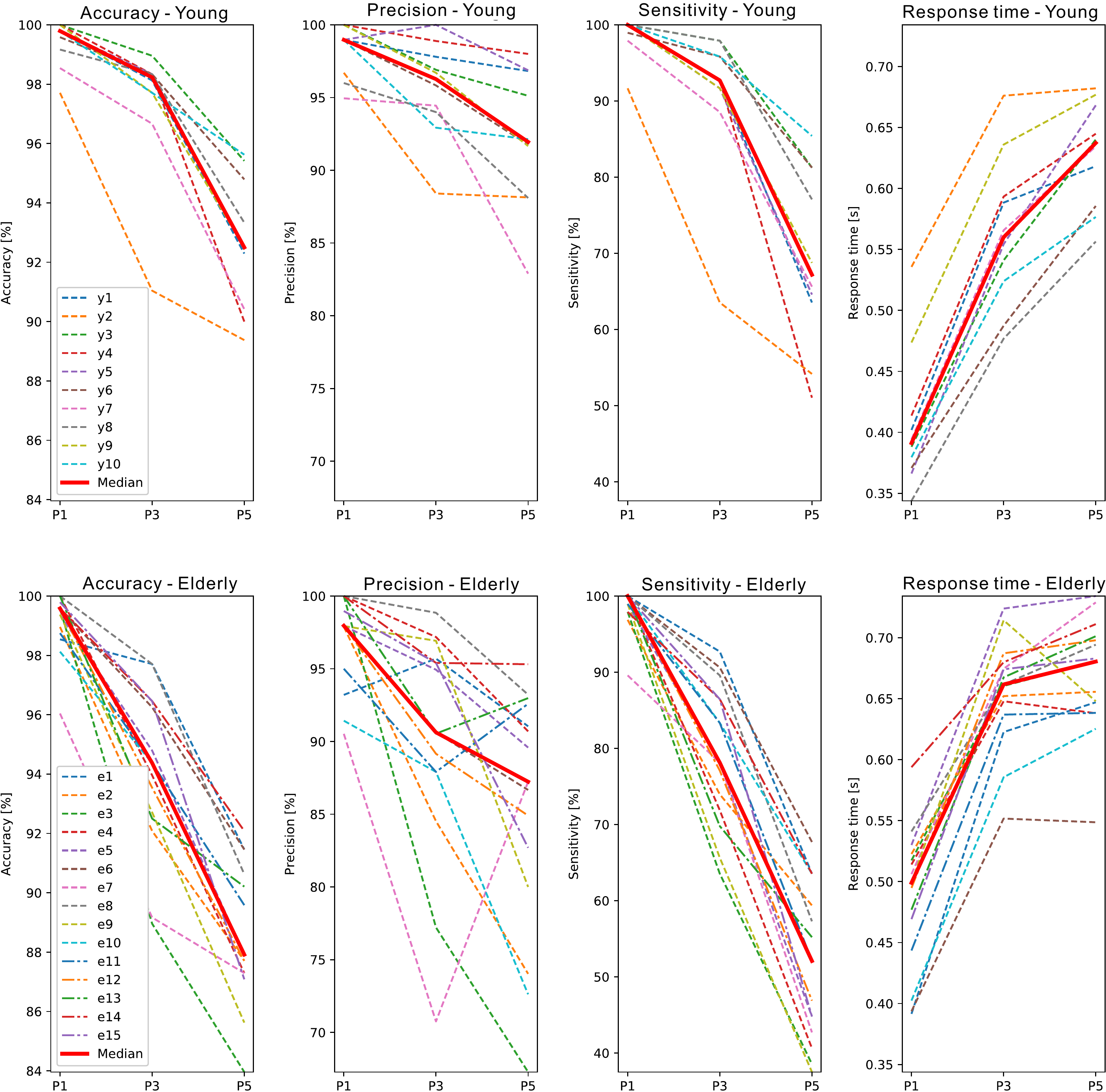}\\
\caption{Individual accuracy, precision, sensitivity, and response time for each stimulus condition.}
\label{fig:acc_place}
\end{figure}

Responses time are presented in Fig.~\ref{fig:acc_place}. 
The median response times of the young participants were 0.39, 0.56, and 0.63 s, 
while those of the elderly participants were 0.50, 0.66, and 0.68 for P1, P3, and P5 conditions, respectively. 
Two types of tendency can be seen: (1) response time gradually increased as the condition changed from P1 to P3, 
and (2) response time saturated at P3 and a similar response time was observed in P5. 
Most young participants exhibited the former tendency, while most elderly participants demonstrated the latter. 
Participant y2 exhibited the latter type. 
The peak response time of participant e9 occurred at P3.

\subsubsection{ERP amplitude and ERP latency}

\begin{figure}[!t]
\centering
\includegraphics[width=10cm]{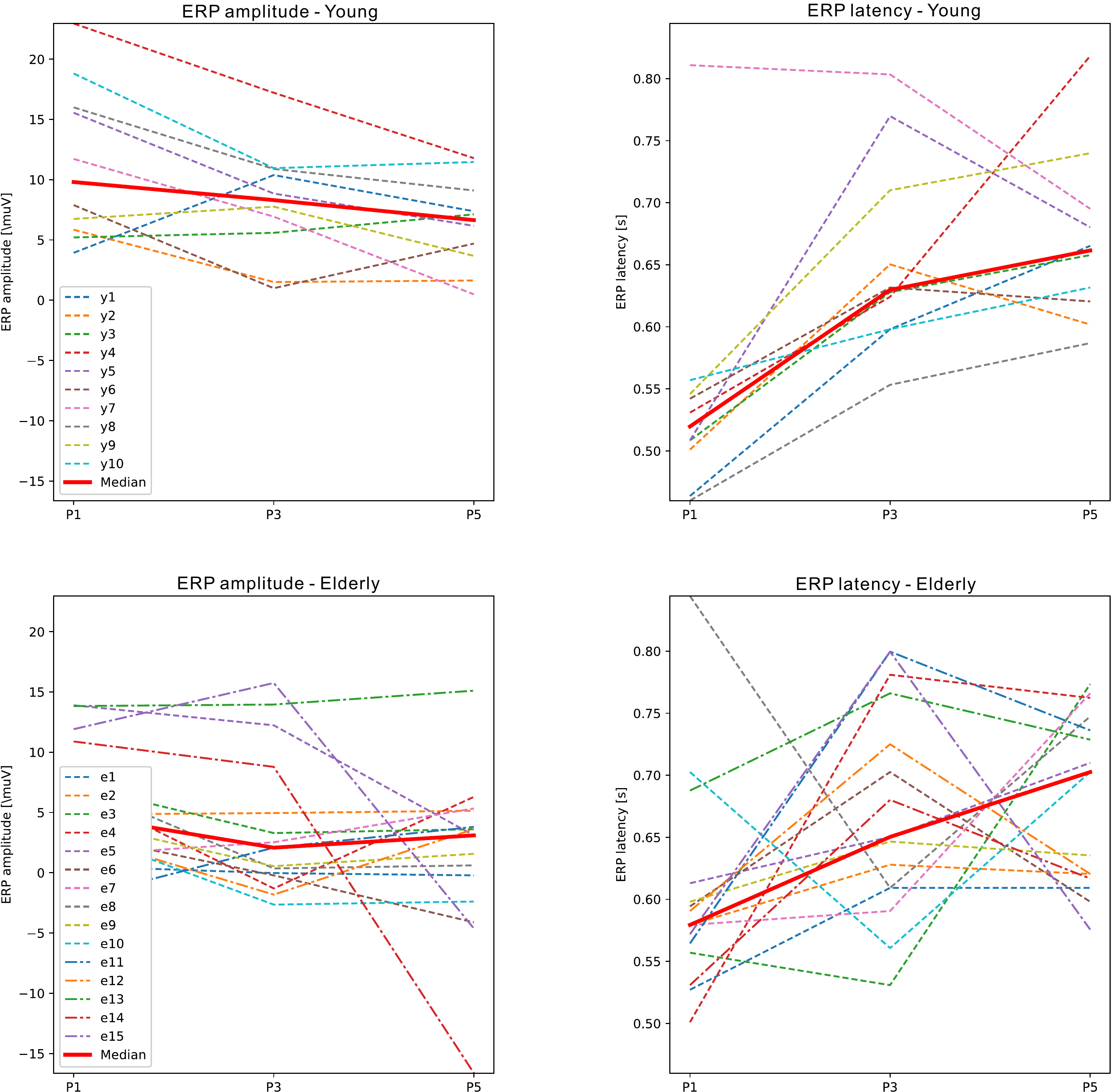}\\
\caption{Individual ERP amplitude and ERP latency for each stimulus condition.}
\label{fig:erp_place}
\end{figure}

ERP amplitude and EPR latency are shown in Fig.~\ref{fig:erp_place}. 
Large differences in median ERP amplitude cannot be seen across stimulus conditions because of its large variance. 
Median EPR amplitude for young participants was 9.8, 8.3, and 6.6 $\mu$V, 
while that of elderly participants was 4.8, 2.1, and 3.1 $\mu$V for P1, P3, and P5 conditions, respectively. 

ERP latency tended to increase as the stimulus condition moved from P1 to P5. 
Median ERP latency for young participants was 0.52, 0.63, and 0.66 s,
while that of elderly participants was 0.58, 0.65, and 0.70 s, respectively. 

\subsection{Statistical tests between stimulus conditions}

The Friedman test was applied to the results shown in 
Fig.~\ref{fig:conf_place}, \ref{fig:acc_place}, and \ref{fig:erp_place}, 
to assess the differences among stimulus conditions for each group. 
The results of the statistical tests are summarized in Table.~\ref{Table:Freeedman}. 
TP differed significantly among all stimulus conditions. 
Significant differences can be seen between P1 and P5 for all indices, 
except for ERP amplitude and ERP latency of young participants, and ERP amplitude for elderly participants. 
P1 and P3 conditions also revealed significant differences between all but TN, FP, and ERP amplitude for young participants, 
and all but ERP amplitude and ERP latency for elderly participants. 
Significant differences between P3 and P5 were confirmed for 
TP, FN, accuracy, precision, sensitivity, and reaction time for young participants, 
and TP, TN, accuracy, and sensitivity for elderly participants. 

\begin{table}[htb]
\centering
\caption{Friedman test.}
\label{Table:Freeedman}
\begin{center}
\begin{tabular}{c|ccc|ccc}
\bhline{2pt}  
Condition 		& ~ 	& Young 	& ~ 	& ~		& Elderly 	& ~ 	\\
~				& P1-P3		& P1-P5		& P3-P5		& P1-P3		& P1-P5		& P3-P5 \\
\hline
TP				& *			& *			& *			& **			& **			& **	\\
TN				& n.s.		& *			& n.s.		& *			& **			& n.s.	\\
FP				& n.s.		& *			& n.s.		& *			& **			& n.s.	\\
FN				& *			& *			& *			& **			& **			& **	\\
Accuracy			& *			& *			& *			& **			& **			& **	\\
Precision		& *			& *			& *			& **			& **			& n.s.	\\
Sensitivity		& *			& *			& *			& *			& *			& *		\\
Reaction time	& *			& *			& *			& *			& *			& n.s.	\\
ERP amplitude	& n.s.		& n.s.		& n.s.		& n.s.		& n.s.		& n.s.	\\
ERP latency		& *			& n.s.		& n.s.		& n.s.		& *			& n.s.	\\
\hline
\end{tabular}

\small\it *:$p<0.05$, **:$p<0.01$, ***:$p<0.001$. 
\end{center}
\end{table}

\clearpage

\subsection{Correlations}

Correlation analysis was applied to results of Fig.~\ref{fig:conf_place}, \ref{fig:acc_place}, and \ref{fig:erp_place}, 
in addition to the profile of participants (Table.~\ref{Table:profile}). 
Correlations among age, TP, TN, FP, and FN are shown in Table.~\ref{Table:age-conf}. 
Significant negative correlations were present between age and TP in P3 and P5 conditions. 
Consistent with the inverted relationship between TN and FN was in Fig.~\ref{fig:conf_place}, 
positive correlations were observed between age and FN in P3 and P5 conditions. 
There were no significant correlations between age and FP nor between age and TN. 

\begin{table}[htb]
\centering
\caption{Correlations among age, TP, TN, FP, and FN.}
\label{Table:age-conf}
\begin{center}
\begin{tabular}{c|ccc|ccc}
\bhline{2pt}  
Condition 	& ~ 	& TP 	& ~ 	& ~		& FN 	& ~ \\
~ 			& P1 	& P3 	& P5 	& P1 	& P3 	& P5  \\
\hline
Age			& n.s.	& -0.43	& -0.58	& n.s.	& 0.43	& 0.58 \\  
\hline \hline
Condition 	& ~ 	& FP 	& ~ 	& ~		& TN 	& ~ \\
~ 			& P1 	& P3 	& P5 	& P1 	& P3 	& P5  \\
\hline
Age			& n.s.	& n.s.	& n.s.	& n.s.	& n.s.	& n.s.	\\  
\hline
\end{tabular}

\small\it Correlation coefficient (value) is listed if the correlation was significant.
\end{center}
\end{table}

Table.~\ref{Table:age-acc} lists correlations among age, accuracy, precision, and sensitivity. 
Significant negative correlations were observed between age and accuracy (P3, P5), 
between age and precision (P5), and between age and sensitivity (P3, P5). 

\begin{table}[htb]
\centering
\caption{Correlations among age, accuracy, precision, and sensitivity.}
\label{Table:age-acc}
\begin{center}
\begin{tabular}{c|ccc|ccc|ccc}
\bhline{2pt}  
Condition 	& ~ 	& Accuracy 	& ~ 	& ~		& Precision & ~ 	& ~		& Sensitivity 	& ~\\
~ 			& P1 	& P3 		& P5 	& P1 	& P3 		& P5  	& P1 	& P3 			& P5\\
\hline
Age			& n.s.	& -0.47		& -0.64	& n.s.	& n.s.		& -0.43	& n.s.	& -0.43			& -0.58 \\  
\hline
\end{tabular}

\small\it Correlations coefficient (value) is listed if the correlation was significant.
\end{center}
\end{table}

Table.~\ref{Table:age-reaction} presents correlations between age and reaction time. 
Correlations were significant for P1 and P3 conditions. 

\begin{table}[htb]
\centering
\caption{Correlations between age and reaction time.}
\label{Table:age-reaction}
\label{Table:}
\begin{center}
\begin{tabular}{c|ccc}
\bhline{2pt}  
Condition 	& ~ 	& Reaction time 	& ~ \\
~ 			& P1 	& P3 				& P5 \\
\hline
Age			& 0.51	& 0.64				& n.s. \\  
\hline 
\end{tabular}

\small\it Correlation coefficient (value) is listed if the correlation was significant.
\end{center}
\end{table}

Table.~\ref{Table:age-ERP} presents correlations among age, ERP amplitude, and ERP latency.
Only age and ERP amplitude in the P1 condition were significantly negatively correlated. 

\begin{table}[htb]
\centering
\caption{Correlations among age, ERP amplitude, and ERP latency.}
\label{Table:age-ERP}
\begin{center}
\begin{tabular}{c|ccc|ccc}
\bhline{2pt}  
Condition 	& ~ 	& ERP amplitude 	& ~ 	& ~		& ERP latency 	& ~ 	\\
~ 			& P1 	& P3 				& P5 	& P1 	& P3 			& P5  	\\
\hline
Age			& -0.51	& n.s.				& n.s.	& n.s.	& n.s.			& n.s. \\  
\hline
\end{tabular}

\small\it Correlation coefficient (value) is listed if the correlation was significant.
\end{center}
\end{table}

Table.~\ref{Table:conf-acc-ERP} summarizes the correlations among TN (P1), FP (P1), 
accuracy (P1), precision (P1), ERP amplitude, and ERP latency.
TN in the P1 condition correlated positively with ERP amplitude in P1 and P3 conditions. 
FP in the P1 condition and ERP amplitude (P1, P3) were negatively correlated. 
There was a significant correlation between accuracy in the P1 condition and ERP amplitude in the P1 condition. 
Precision in the P1 condition also correlated positively with ERP amplitude in P1 and P3 conditions. 

\begin{table}[htb]
\centering
\caption{Correlations among TN (P1), FP (P1), accuracy (P1), precision (P1), ERP amplitude, and ERP latency.}
\label{Table:conf-acc-ERP}
\begin{center}
\begin{tabular}{c|ccc|ccc}
\bhline{2pt}  
Condition 		& ~ 	& ERP amplitude 	& ~ 	& ~		& ERP latency 	& ~ 	\\
~ 				& P1 	& P3 				& P5 	& P1 	& P3 			& P5  	\\
\hline
TN (P1)			& 0.51 	& 0.44 				& n.s.	& n.s.	& n.s.			& n.s.	\\
FP (P1)			& -0.51 & -0.44 			& n.s.	& n.s.	& n.s.			& n.s.	\\
Accuracy (P1)	& 0.45 	& n.s.				& n.s.	& n.s.	& n.s.			& n.s.	\\
Precision (P1)	& 0.51 	& 0.43 				& n.s.	& n.s.	& n.s.			& n.s.	\\
\hline
\end{tabular}

\small\it Correlation coefficient (value) is listed if the correlation was significant. 
Data not shown in this table, such as TN(P3) and TN(P5) conditions in addition to TP, FN, and sensitivity, 
did not exhibit any significant correlations with either ERP amplitude or ERP latency. 
\end{center}
\end{table}

\clearpage

\section{Discussion}

The OSVS task was proposed and evaluated to develop a low-cost, rapid aptitude test for elderly drivers. 
During the task, the number of stimuli was switched to 
1, 3, and 5 in P1, P3, and P5 conditions, respectively. 
Reactions were measured by button-press and EEG. 
As a result, poor performers were confirmed in TP, FN, accuracy, precision, and response time. 
Furthermore, TP, FN, accuracy, precision, and response time in the P3 condition were correlated with age. 
These results imply that performance measured by button-press responses 
may be indicative of driving performance among elderly drivers. 
The prototype test device without EEG was cheap, because it consisted of only a computer and button-response box. 
In addition, the task duration was somewhat extended so as to conduct a rigorous psychological experiment, 
but the duration could be shortened. 
Therefore, the OSVS task may be suitable as a low-cost, rapid aptitude test for elderly drivers.

This study showed that in the P3 condition, some performance measures correlated with age. 
The P3 condition of the OSVS task is a type of selective attention task. 
As mentioned in the introduction, 
age is related to decline in visual cognitive functions, attention, and UFOV. 
Therefore, the results of this study are consistent with those of previous studies. 
Some young participants also exhibited poor performance in the P3 condition. 
The results imply that some young drivers 
may have poor attention-related performance, 
which could be detected by the OSVS task.  
Indeed, young drivers are relatively likely to cause serious driving accidents. 
Young drivers often believe that their driving performance is high; as such,
they tend to overestimate their own skills. 
In contrast, a study showed that young drivers are not optimistic \cite{de2011young}. 
Young drivers tend to cause accidents in some situations, 
such as late at night \cite{williams2003teenage, braitman2008crashes}. 
Therefore, aptitude tests or training for young drivers could 
reduce such traffic accidents. 
That is, the OSVS task may contribute to reducing traffic accidents 
among younger drivers as well as among elderly drivers. 

Although the OSVS test was intended as an aptitude test that is used when obtaining or renewing driver's licences, 
the test could be used in other applications. 
For example, it could be employed as a self-assessment when introducing 
a driving assistance system \cite{bengler2014three} or automated driving in future 
\cite{milakis2017policy, taiebat2018review}. 
Furthermore, it would be interesting if the training effect of the OSVS test was clarified. 
In this respect, training methods for elderly drivers have been proposed. 
For example, a trail making game system has been proposed for training \cite{hiraoka2016cognitive}.
Additionally, a study proposed a video-based trail making test for young drivers \cite{sasaki2017development}. 
OSVS has the potential to be extended as a training tool for young and elderly drivers. 

EEG amplitude correlated with responses in P1, but not those in P3 or P5. 
That is, EEG amplitude was not associated with selective attention. 
The OSVS task with the P3 or P5 conditions is a type of selective attention task,
while the P1 condition represents a sustained attention task. 
Indeed, the oddball task is used for measuring sustained attention \cite{hwang2019segregating}.

In future studies, the correlation 
between OSVS scores and the number of traffic accidents should be identified. 
Previous studies reported that the correlation between 
cognitive function scores and traffic accidents was low \cite{hills1974vision, burg1968vision}. 
In contrast, UFOV test outcomes were correlated with the number of traffic accidents, as mentioned in the introduction. 
OSVS scores, which can be obtained without the need for expensive devices, were not compared with the number of traffic accidents. 
Accordingly, OSVS task scores and driving simulator tests should be compared. 
Such a comparison is required to determine the threshold of OSVS scores for the aptitude test. 
Additionally, a shortened OSVS task should be evaluated to establish a low-cost, rapid aptitude test.

\section{Conclusion}

This study developed a low-cost, rapid aptitude test for elderly drivers. 
Since psychological tests are more economical than are DS tests, 
this study focused on a simple psychological test that required only a computer and a response button. 
This study proposed a new aptitude test that combined a sequential visual search task with an oddball task. 
The number of stimuli varied in three conditions to clarify individual differences in performance. 
Poor performers were confirmed in terms of TP, FN, accuracy, precision, and response time. 
Further, those scores were correlated with age when three stimuli were simultaneously presented. 
The proposed task may contribute to developing 
a low-cost, rapid aptitude test for elderly drivers.

\section{Acknowledgements}

This study was supported in part by the Mitsui Sumitomo Insurance Welfare Foundation.

\bibliography{mybibfile}

\end{document}